\documentclass[12pt]{article}

        \usepackage{amsfonts}
        \usepackage{amssymb}

\def\be{\begin{eqnarray}}
\def\ee{\end{eqnarray}}
\def\bee{\begin{eqnarray*}}
\def\eee{\end{eqnarray*}}

\newtheorem{thm}{Theorem}
\newtheorem{cor}[thm]{Corollary}
\newtheorem{lemma}[thm]{Lemma}
\newtheorem{conj}[thm]{Conjecture}

        \def\half{{\textstyle \frac{1}{2}}}
        \def\tr{\hbox{Tr}}
        
      \def\Hil{{\cal H}}

\def\holv{{\rm Holv}}

\def\bra{\langle}
\def\ket{\rangle}

\def\ot{\otimes}

\def\half{{\textstyle \frac{1}{2}}}

\def\Tr{{\rm Tr}}

          \title{
Maximization of capacity and $l_p$ norms for some
product channels.}

        \author{Christopher King \\ Department of
        Mathematics \\ Northeastern University \\
 Boston, MA 02115  \\ {\normalsize king@neu.edu}
}
\begin{document}

\maketitle

\begin{abstract}
It is conjectured that the Holevo capacity of a product channel
$\Omega \otimes \Phi$
is achieved when product states are used as input.
Amosov, Holevo and Werner have also conjectured that the
maximal $l_p$ norm of a product channel is achieved
with product input states. In this paper
we establish both of these conjectures
in the case that $\Omega$ is arbitrary and $\Phi$ is
a CQ or QC channel (as defined by Holevo). We also establish
the Amosov, Holevo and Werner conjecture when $\Omega$ is
arbitrary and either
$\Phi$ is a qubit channel and $p=2$, or
$\Phi$ is a unital qubit channel and $p$ is integer.
Our proofs involve a new conjecture for the norm of an output
state of the half-noisy channel $I \otimes \Phi$,
when $\Phi$ is a qubit channel. We show that this conjecture
in some cases also implies additivity of the Holevo capacity. 
\end{abstract}

\pagebreak

%\tableofcontents

%\bigskip

\section{Introduction}
A quantum channel is the mathematical description of a device which
stores and transmits quantum states. Much work has been devoted to the
study of particular quantum channels with highly non-classical
properties, and also to general questions such as the information
capacity of classes of channels.  In this paper we will consider some
problems of the second type, concerning additivity and multiplicativity properties 
that are believed to hold for all product channels.

The basic components of a quantum channel are a Hilbert
space $\Hil$ and a noise operator $\Phi$. The quantum states are
positive operators on $\Hil$, with trace equal to one. The noise
operator $\Phi$ is a completely positive, trace-preserving map which
acts on the set of states. Positivity means that $\Phi$ is
a positive operator on $B(\Hil)$ (the algebra of bounded
operators on $\Hil$). Complete positivity
means that the map $I \otimes \Phi$ is also a positive operator
on $B({\bf C}^K \otimes \Hil)$ for every $K$.

When the channel $(\Hil, \Phi)$ is used to store or transmit information,
it is assumed that the information is encoded as a state on
the product space ${\Hil}^{\otimes n}$ for some $n$, and that the noise
acts on this state through the product operator ${\Phi}^{\otimes n}$, 
thereby mimicking the action of
a memoryless channel in classical information theory. The basic properties
of such quantum memoryless channels have been studied by many
authors \cite{BS}, \cite{Fu1}, \cite{Hv1}, \cite{Hv2}, \cite{SW1}. One outstanding problem is to
determine the ultimate rate at which classical information can be
transmitted through this channel, when no prior entanglement is available
between sender and receiver. The protocol that achieves this
capacity may require messages to
be encoded using entangled states and/or decoded using collective measurements.
It is conjectured that this ultimate capacity is
given by the well-known Holevo bound \cite{Hv1}
\be\label{Hol}
C_{\holv}(\Phi) = \sup_{\pi, \, \rho} \bigg[S \big(\sum {\pi}_i \Phi({\rho}_{i})\big)
- \sum {\pi}_i S(\Phi({\rho}_{i})) \bigg],
\ee
where $S(\rho) = - \Tr \rho \log \rho$ is the von Neumann entropy,
and the $\sup$ runs over all probability distributions $\{{\pi}_{i}\}$
and collections of states $\{{\rho}_{i}\}$ on $\Hil$.
This capacity conjecture is equivalent to the statement that there is
no benefit gained when entangled states are used to encode messages for
transmission through a quantum channel.
As shown by Holevo \cite{Hv1} and Schumacher-Westmoreland \cite{SW1},
the ultimate rate for information transmission using non-entangled
coding states is exactly $C_{\holv}$. Thus the capacity conjecture is
implied by the {\it additivity conjecture} for $C_{\holv}$, which states
that for any channels $\Omega$ and $\Phi$
\be\label{add.Hol}
C_{\holv}(\Omega \otimes \Phi) =  C_{\holv}(\Omega) + C_{\holv}(\Phi)
\ee

Although the equality (\ref{add.Hol}) has been shown in some special cases
\cite{AHW}, \cite{BFMP}, \cite{Hv2}, \cite{KR}, \cite{SW1}, 
it remains a challenging problem to prove this result for a general pair of
channels $(\Omega,\Phi)$. Amosov, Holevo and Werner introduced a related
conjecture, concerning the noncommutative $l_p$ norm 
of output states from a product channel \cite{AHW} (this norm is
defined below). In this paper we report progress toward establishing
these conjectures for some special product channels, namely the cases
when $\Omega$ is arbitrary and either (i) $\Phi$ is a CQ or QC channel
(these are defined below), or (ii) $\Phi$ is a qubit channel. In the first case
we establish both conjectures. In the second case we establish the 
Amosov, Holevo and Werner conjecture for integer values of $p$.
A principal ingredient in our proof in the second case is a
new bound concerning the $l_p$ norm of the output from  a ``half-noisy'' channel
$I \otimes \Phi$, for integer values of $p$. 
We conjecture that this bound holds for all $p \geq 1$, and we show that in some cases this
conjecture implies additivity of the Holevo bound (\ref{add.Hol}).

\medskip

The paper is organised as follows. Section 2 contains a precise statement
of the results, and the conjectured bound for
half-noisy channels. In section 3 we review the
relation of relative entropy and the Holevo bound. In sections 4 and 5 we
prove the results for CQ and QC
channels. Then in section 6 we prove the results for qubit channels, and in
section 7 we prove the Corollaries of our new conjecture. In section
8 we give a summary and overview of the results in the paper. Finally
the Appendix contains a proof by Lieb and Ruskai of a special case of
the conjecture.

\section{Statement of results}
The noncommutative $l_p$ norm of a matrix $A$ is defined by
\be\label{def:l_p}
|| A ||_{p} = \big( \Tr |A|^{p} \big)^{1 \over p} =
\bigg[ \Tr (A^{*} A)^{p \over 2} \bigg]^{1 \over p}
\ee
The corresponding maximal $l_p$ norm for a positive map $\Phi$ on $B(\Hil)$ is
\be\label{def:norm}
{\nu}_{p}(\Phi) = \sup_{\rho}  || \Phi(\rho) ||_{p}
\ee
where the $\sup$ runs over states in $\Hil$ (this quantity was introduced
in \cite{AHW}, where it was called the
`maximal output purity' of the channel). It is always true that for any maps
$\Omega$ and $\Phi$, and any $p \geq 1$
\be\label{ineq}
{\nu}_{p}(\Omega \otimes \Phi) \geq {\nu}_{p}(\Omega) \, {\nu}_{p}(\Phi)
\ee
The multiplicativity
conjecture of \cite{AHW} states that for any completely positive trace-preserving maps
$\Omega$ and $\Phi$, and for all $p \geq 1$,
\be\label{AHWconj}
{\nu}_{p}(\Omega \otimes \Phi) = {\nu}_{p}(\Omega) \, {\nu}_{p}(\Phi)
\ee

Equality always holds in (\ref{AHWconj}) for $p=1$.
It has been shown in several different ways that (\ref{AHWconj}) holds
for all $p \geq 1$ and all $\Omega$ when $\Phi = I$ \cite{AHW}, \cite{Fu2}, \cite{SW2}. 
Recently, it has been shown that (\ref{AHWconj}) holds when both
$\Omega$ and $\Phi$ are depolarizing channels, and $p$ is integer \cite{AH}.
In this paper we provide some further examples where it holds.

The first case we consider involves the CQ and QC channels introduced by 
Holevo \cite{Hv2}, so
we recall their definitions now. Let $\{X_b\}$ be a POVM on $\Hil$ (so
$X_b \geq 0$ and $\sum X_b = I$) and let $\{Q_b\}$ be any collection of states.
Then we can define a channel $\Phi$ by the formula
\be\label{gen.CQC}
\Phi(\rho) = \sum \Tr (\rho \, X_b ) \, Q_b
\ee
Holevo considered two special cases of (\ref{gen.CQC}). First, if
$\{X_b = |e_b \ket \bra e_b |\}$ are projections onto
an orthonormal basis $\{|e_b \ket\}$ in $\Hil$, then (\ref{gen.CQC}) 
is called a CQ channel. Second, if 
$\{Q_b = |e_b \ket \bra e_b |\}$, then (\ref{gen.CQC}) 
is called a QC channel.
Holevo proved the additivity result (\ref{add.Hol}) when
$\Omega = \Phi$ is  either a  CQ or QC channel. Our first result
generalises this by allowing an arbitrary channel $\Omega$.

\begin{thm}\label{thm1}
Let $\Phi$ be a {\rm CQ} or {\rm QC}  channel.
Then for any completely positive trace-preserving map
$\Omega$, $l_p$-multiplicativity (\ref{AHWconj}) holds for all $p \geq 1$,
and Holevo additivity (\ref{add.Hol}) holds.
\end{thm}
\bigskip

For our second set of results we restrict to channels on a two-dimensional
Hilbert space.
For brevity of notation we will say that $\Phi$ is a {\it qubit map} if
it is a completely positive trace-preserving map on
${\cal B}({\bf C}^2)$. 

\begin{thm}\label{thm2}
Let $\Phi$ be a qubit channel. Then the equality (\ref{AHWconj}) holds for $p = 2$, that is
${\nu}_{2}(\Omega \otimes \Phi) = {\nu}_{2}(\Omega) \,\, {\nu}_{2}(\Phi)$
for all channels $\Omega$.
\end{thm}
\bigskip

In order to state the next result
we need to recall the classification of qubit maps.
Any qubit map $\Phi$ can be represented by a real $4 \times 4$ matrix
with respect to the basis $I, {\sigma}_{1}, {\sigma}_{2}, {\sigma}_{3}$,
where ${\sigma}_{i}$ are the Pauli matrices. In \cite{KR} it was explained
that by using independent unitary transformations in
its domain and range, this matrix can be put into the 
following form:
\be\label{Phi}
\Phi = \left(\matrix{1 & 0 & 0 & 0 \cr
t_{1} & {\lambda}_{1} & 0 & 0 \cr
t_{2} & 0 & {\lambda}_{2} & 0 \cr
t_{3} & 0 & 0 & {\lambda}_{3} \cr} \right)
\ee
This form makes it easy to see how $\Phi$ acts on the Bloch
sphere. The sphere is first compressed to an ellipsoid
with semi-major axes $|{\lambda}_{1}|, |{\lambda}_{2}|, |{\lambda}_{3}|$,
and is then translated by the vector ${\bf t} = (t_1, t_2, t_3)$.
There are constraints on the allowed values of these six parameters
(coming from the requirements that $\Phi$ be completely positive and
trace-preserving), and
these constraints have been fully worked out in \cite{RSW}. 
If $t_i = 0$ for $i = 1,2,3$ then $\Phi (I) = I$, in which case $\Phi$ is a
{\it unital} qubit map.

Our next result requires a slightly stronger condition
on the map $\Phi$, which we now state in terms of these parameters: 
\be\label{condition}
\quad {\rm if} \quad |{\lambda}_{i}| < |{\lambda}_{j}| < |{\lambda}_{k}|
\quad {\rm then} \quad t_{i} t_{j} = 0
\ee
This condition can be stated in words as follows: the ellipsoid may be translated
only in directions lying in the two planes that are perpendicular to
its two smaller axes (if any two axes have equal length,
there is no restriction).

\begin{thm}\label{thm3}
Let $\Phi$ be a qubit channel
satisfying the condition (\ref{condition}). Then
$l_p$-multiplicativity (\ref{AHWconj}) holds for all integer $p$,
that is ${\nu}_{p}(\Omega \otimes \Phi) = {\nu}_{p}(\Omega) \,\, {\nu}_{p}(\Phi)$
for all channels $\Omega$ and all integers $p$.
\end{thm}
\bigskip

The proofs of Theorem \ref{thm2} and Theorem \ref{thm3} 
make use of a bound for the $l_p$ norm of the output state
from the half-noisy channel $I \otimes \Phi$. We believe that
this bound holds for all $p \geq 1$, however we can prove it only for the
cases listed in the Theorems. So we state the general bound as a
conjecture.

\bigskip
\begin{conj}\label{conj1}
Let $\Phi$ be a qubit channel, and let $M \geq 0$ be a $2 K \times 2 K$
matrix. Write $M$ in the form
\be\label{def:M}
M = \left(\matrix{X & Y \cr Y^{*} & Z }\right),
\ee
where $X$, $Y$ and $Z$ are $K \times K$ matrices. Then for all $p \geq 1$
\be\label{new}
|| (I \otimes \Phi) (M) ||_{p} \leq {\nu}_{p}(\Phi) \,\, \big(
|| X ||_{p} + || Z ||_{p} \big)
\ee
\end{conj}
\bigskip

This conjecture has several important consequences, which we 
list in the next three Corollaries. In particular, the first
Corollary  shows that  Conjecture \ref{conj1}
implies Theorems \ref{thm2} and \ref{thm3}.

\bigskip
\begin{cor}\label{cor1}
Let $\Phi$ be a qubit channel, and suppose that (\ref{new}) holds for all
positive $2K \times 2K$  matrices $M$, for some $p \geq 1$.
Then for any completely positive map $\Omega$ on $B({\bf C}^K)$, 
$l_p$-multiplicativity (\ref{AHWconj}) holds for the same value of $p$. 
\end{cor}

\bigskip
In Section 5 we will prove that (\ref{new}) holds for all qubit maps $\Phi$ when $p=2$,
and also for the cases listed in Theorem \ref{thm3}. Combining this with
Corollary \ref{cor1} will prove Theorems \ref{thm2} and
 \ref{thm3}.

\bigskip
Our next result concerns
the additivity of minimal entropy. The minimal entropy of a 
completely positive trace-preserving map $\Phi$ is defined by
\be\label{def:S_min}
S_{\rm min}(\Phi) = \inf_{\rho} S(\Phi(\rho))
\ee
The additivity of minimal entropy is the statement that
\be\label{add.Smin}
S_{\rm min}(\Omega \otimes \Phi) = S_{\rm min}(\Omega) + S_{\rm min}(\Phi)
\ee

\bigskip
\begin{cor}\label{cor2}
Let $\Phi$ be a qubit channel, and suppose that (\ref{new}) holds for all
positive $2K \times 2K$  matrices $M$,
and for all $p \in [1,s)$ for some  $s > 1$.
Then for any completely positive map $\Omega$ on $B({\bf C}^K)$, 
additivity of minimal entropy (\ref{add.Smin}) holds. 
\end{cor}

\bigskip
For our last corollary,  recall that a map $\Phi$ is unital if
$\Phi (I) = I$, which means roughly that $\Phi$ leaves unchanged the
``noisiest'' state through the channel.

\bigskip
\begin{cor}\label{cor3}
Let $\Phi$ be a unital qubit channel, and suppose that (\ref{new}) holds for all
positive $2K \times 2K$  matrices $M$, and for all $p \in [1,s)$ for some  $s > 1$.
Then for any completely positive trace-preserving map $\Omega$ on $B({\bf C}^K)$, 
Holevo additivity (\ref{add.Hol}) holds.
\end{cor}

\bigskip
\noindent{\it Remarks}.
\medskip
\par\noindent {\bf 1)}
There are two special cases where it is easy to verify Conjecture \ref{conj1}.
First, if $M$ is a one-dimensional projection then the right side of
(\ref{new}) becomes ${\nu}_{p}(\Phi) \,\, \Tr (M)$, and then the result follows
immediately from the definition (\ref{def:norm}). Second, suppose that $\Phi$ is
the identity map, so ${\nu}_{p}(\Phi) = 1$.  Define the projections
\be
P_{0} = \left(\matrix{I & 0 \cr 0 & 0 }\right), \quad\quad
P_{1} = \left(\matrix{0 & 0 \cr 0 & I }\right)
\ee
Then convexity of the $l_p$ norm for $p \geq 1$ implies that 
\be
|| M ||_{p} = || M^{1/2} (P_{0} + P_{1}) M^{1/2} ||_{p}
\leq || M^{1/2} P_{0} M^{1/2} ||_{p} + || M^{1/2}  P_{1} M^{1/2} ||_{p}
\ee
Furthermore for any matrix $A$, the matrices $A A^{*}$ and $A^{*} A$ have the same
spectrum, so we deduce that
\be
|| M ||_{p} \leq || P_{0} M P_{0} ||_{p} +
|| P_{1} M P_{1} ||_{p} = || X ||_{p} + || Z ||_{p}
\ee
(this derivation is a special case of a more general result for
POVM's which is described in \cite{Fu3}).

\medskip
\par\noindent {\bf 2)}
Lieb and Ruskai have recently established Conjecture \ref{conj1}, 
eq. (\ref{new}) for a
depolarizing channel in the special case  $X = Z$, for all $p \geq 1$. 
Recall that the depolarizing channel
is described by the parameter values
${\lambda}_{1} = {\lambda}_{2} = {\lambda}_{3} = \lambda$, and $t_1 = t_2 = t_3 = 0$,
so that in this case the bound (\ref{new}) becomes
\be\label{LR}
\bigg{|}\bigg{|} \left(\matrix{X & \lambda Y \cr \lambda Y^{*} & X }\right) \bigg{|}\bigg{|}_{p}
\leq {\nu}_{p}(\Phi) \,\, \big( 2 \,
|| X ||_{p} \big)
\ee
where
\be
{\nu}_{p}(\Phi) = \bigg[ \bigg({1 + \lambda \over 2}\bigg)^{p} + 
\bigg({1 - \lambda \over 2}\bigg)^{p} \bigg]^{1/p}
\ee
Their proof appears as an Appendix to this paper.

\medskip
\par\noindent {\bf 3)} 
Theorem \ref{thm3} was proved in \cite{KR} for unital maps
in the case $p = \infty$,
and our proof here extends this result to all integer values of $p$
(and to a larger class of maps).
The class of qubit maps which satisfy (\ref{condition}) includes all unital qubit maps and
many non-unital maps. In particular, our proof applies
to any extreme point in the set of qubit maps (this refers to recent work
in \cite{RSW}, and we discuss it more fully in section 3).

\medskip
\par\noindent {\bf 4)} 
To prove Corollaries \ref{cor2} and \ref{cor3} we need 
only the derivative of (\ref{new}) at $p=1$, which we
now state as a separate bound.
Assume that $M$ has the form (\ref{def:M}) with $\Tr (M) = 1$, and define
the states
\be\label{def:xi,zeta}
{\xi} = {1 \over \Tr X }\,\, X, \quad\quad
{\zeta} = {1 \over \Tr Z }\,\, Z
\ee
Then taking the derivative of (\ref{new}) at $p=1$ gives
\be\label{new'}
S\big( (I \otimes \Phi) (M) \big) \geq S_{\rm min}(\Phi) + \Tr (X) S(\xi)
+ \Tr (Z) S(\zeta)
\ee

\medskip
\par\noindent {\bf 5)}
When $\Omega$ and $\Phi$ are both unital qubit maps, the additivity result
(\ref{add.Hol}) follows immediately from the additivity of minimal entropy
(\ref{add.Smin}),
as was discussed in \cite{KR}. This is also true if $\Omega = {\Phi}_{1}
\otimes \cdots \otimes {\Phi}_{n}$ is a product of unital qubit maps.
Additivity of Holevo capacity (\ref{add.Hol}) for the `half-noisy' case
$\Phi = I$ was proved by Schumacher and Westmoreland \cite{SW2}, and their
analysis underlies our proof of Corollary \ref{cor3}.
\medskip

\section{Relative entropy and the Holevo bound}
The Holevo bound (\ref{Hol}) can be re-expressed in terms of
relative entropy in several ways (see for example the discussion in
\cite{KR2}). Here we will follow the approach of Ohya, Petz and Watanabe \cite{OPW}
and Schumacher and Westmoreland \cite{SW2}, who express (\ref{Hol}) as an
optimization of relative entropy.

Let $\Phi$ be a channel,
and let ${\cal E} = \{{\pi}_{i}, {\rho}_{i}\}$ be an ensemble of input
states for the channel. Define
\be\label{def:chi}
\chi (\Phi; {\cal E}) = S \bigg(\sum {\pi}_i \Phi ({\rho}_{i}) \bigg)
- \sum {\pi}_i S( \Phi({\rho}_{i}))
\ee
Following the notation of \cite{SW2}, the Holevo capacity of the channel is
denoted
\be
{\chi}^{*}(\Phi) = C_{\holv}(\Phi) = \sup_{{\cal E}} \chi (\Phi; {\cal E})
\ee
As shown in \cite{SW2} there is an ensemble which achieves this supremum.
The ensemble may not be unique, however its average input state is unique.
We let ${\rho}^{*} = \sum {\pi}_i {\rho}_{i}$ denote this optimal average
input state.

The relative entropy of a state $\omega$ with respect to a state $\rho$ is
defined by
\be\label{def:S-rel}
S(\omega \,|\, \rho) = \Tr \, \omega \, ( \log \omega - \log \rho)
\ee
Relative entropy is non-negative: $S(\omega \,|\, \rho) \geq 0$,
with equality if and only if $\omega = \rho$. 
There is a useful characterization of the capacity
${\chi}^{*}(\Phi)$ in terms of relative entropy, namely
\be\label{OPWthm}
{\chi}^{*}(\Phi) = \inf_{\rho}\, \sup_{\omega} \, 
S \big(\Phi(\omega) \,|\, \Phi(\rho) \big)
\ee
This result was derived in \cite{OPW} and also in \cite{SW2}.
For our purposes it is convenient to restate it as follows:
\bigskip \noindent
for any state $\rho$,
\be\label{SWthm}
{\chi}^{*}(\Phi) \leq  \sup_{\omega} 
S( \Phi(\omega) \,|\, \Phi({\rho})) 
\ee
and equality holds in (\ref{SWthm}) if and only if 
$\rho = {\rho}^{*}$.
\bigskip

Our goal is the additivity result (\ref{add.Hol}). 
By restricting to product states it is clear that
\be
{\chi}^{*}(\Omega) + {\chi}^{*}(\Phi)  \leq {\chi}^{*}(\Omega \otimes \Phi)
\ee
So to establish (\ref{add.Hol}) it is sufficient to prove the bound
\be\label{Hol.bound}
{\chi}^{*}(\Omega \otimes \Phi) \leq {\chi}^{*}(\Omega) + {\chi}^{*}(\Phi)
\ee
For a channel $\Phi$, denote the optimal average {\it output}
state by
\be\label{def:opt.out}
{\rho}_{\Phi} := \Phi({\rho}^{*})
\ee
Then  (\ref{SWthm}) implies that
\be
{\chi}^{*}(\Omega \otimes \Phi) \leq 
\sup_{\tau} S \bigg( (\Omega \otimes \Phi) (\tau) \,|\, {\rho}_{\Omega} \otimes 
{\rho}_{\Phi} \bigg) 
\ee
Therefore in order to prove (\ref{Hol.bound}), and hence (\ref{add.Hol}),
it is sufficient to show that for any state $\tau$,
\be\label{suff1}
S \bigg( (\Omega \otimes \Phi) (\tau) \,|\, {\rho}_{\Omega} \otimes 
{\rho}_{\Phi} \bigg) \leq {\chi}^{*}(\Omega) + {\chi}^{*}(\Phi)
\ee

\section{Proof for CQ channel}
Let $\Phi$ be a CQ channel on $B({\bf C}^N)$, so that
\be
\Phi(\rho) = \sum \Tr (\rho X_{b}) Q_{b},
\ee
where $\{X_{b}\}$ are one-dimensional orthogonal projections.
It follows that for all $b = 1, \dots, N$,
\be\label{getQ}
Q_{b} = \Phi( X_{b})
\ee
Let $\Omega$ be a completely positive map on $B({\bf C}^K)$.
Then for any state $\tau$ in $B({\bf C}^K \otimes {\bf C}^N)$,
\be\label{CQ1}
( \Omega \otimes \Phi) (\tau) = \sum \Omega \bigg( 
{\Tr}_{2} \big( (I \otimes X_{b} ) \, \tau \big) \bigg) \otimes Q_{b}
\ee
where ${\Tr}_{2}$ is the trace over the second factor. For each
$b = 1, \dots, N$ let
\be\label{def:n_b}
n_{b} = {\Tr}  \big( (I \otimes X_{b} ) \, \tau \big),
\ee 
and define the state
\be\label{def:tau_b}
{\tau}_{b} = {1 \over n_{b}}\,\,
{\Tr}_{2} \big( (I \otimes X_{b} ) \, \tau \big)
\ee
Then (\ref{CQ1}) can be written
\be\label{CQ2}
( \Omega \otimes \Phi) (\tau) = \sum n_{b} \, \Omega ({\tau}_{b})
\otimes Q_{b} =
\sum n_{b} \, 
\Omega ({\tau}_{b}) \otimes \Phi(X_{b})
\ee
where in the second equality we used (\ref{getQ}).

Turning first to the $l_p$ norm result, it follows from
(\ref{CQ2}) and the definition (\ref{def:norm}) that
\be
|| ( \Omega \otimes \Phi) (\tau) ||_{p} \leq
\sum n_{b} \, {\nu}_{p}(\Omega) \, {\nu}_{p} (\Phi) =
{\nu}_{p}(\Omega) \, {\nu}_{p} (\Phi)
\ee
and this proves (\ref{AHWconj}).

Turning next to the channel capacity result, we will prove that
(\ref{suff1}) holds. Indeed (\ref{CQ2}) implies that
\be
S \bigg( (\Omega \otimes \Phi) (\tau) \,|\, {\rho}_{\Omega} \otimes 
{\rho}_{\Phi} \bigg) \leq
\sum n_{b} \, \bigg[
S \bigg( \Omega ({\tau}_{b}) \,|\, {\rho}_{\Omega} \bigg)
+ S \bigg( \Phi(X_{b}) \,|\, {\rho}_{\Phi} \bigg)\, \bigg]
\ee
where we used the additivity of relative entropy for product states.
Now (\ref{OPWthm}) implies
\be
S \bigg( \Omega ({\tau}_{b}) \,|\, {\rho}_{\Omega} \bigg) \leq
{\chi}^{*}(\Omega),
\quad\quad
S \bigg( \Phi(X_{b}) \,|\, {\rho}_{\Phi} \bigg) \leq
{\chi}^{*}(\Phi)
\ee
which proves the result.

\section{Proof for QC channel}
Let $\Phi$ be a QC channel, so that
\be\label{defQC}
\Phi(\rho) = \sum \Tr (\rho X_{b}) Q_{b},
\ee
where $\{Q_{b}\}$ are one-dimensional orthogonal projections.
For any state $\tau$,
\be\label{eqn2}
(\Omega \otimes \Phi) (\tau) & = & \sum \Omega \big({\Tr}_{2} 
(I \otimes X_{b})\tau  \big) \otimes Q_{b} \\ \nonumber
& = & \sum n_{b} \, \Omega ({\tau}_{b}) \otimes Q_{b}
\ee
where we use the definitions (\ref{def:n_b}) and (\ref{def:tau_b}). Now
define
\be
\theta = {\Tr}_{1} (\tau),
\ee
then it follows that
\be
n_{b} = \Tr ( \theta \, X_{b} )
\ee
and (\ref{eqn2}) can be written as
\be\label{eqn3}
(\Omega \otimes \Phi) (\tau) = \sum \Omega ({\tau}_{b}) \otimes 
\big( \Tr ( \theta \, X_{b} ) \, Q_{b} \big)
\ee

\medskip
First we prove the bound for the $l_p$ norm. Using the
fact that $\{Q_{b}\}$ are orthogonal projections, we get
\be\label{eqn4}
\Tr | (\Omega \otimes \Phi) (\tau) |^{p}  = 
\sum \Tr |  \Omega({\tau}_{b}) |^{p} \,\, 
\big( \Tr ( \theta \, X_{b} )\big)^{p}
\ee
The definition of the $l_p$ norm implies that for any positive matrix $A$,
\be
|| \Omega (A) ||_{p} \leq {\nu}_{p}(\Omega) \,\, \Tr (A)
\ee
and hence (\ref{eqn4}) implies that
\be\label{eqn5}
\Tr | (\Omega \otimes \Phi) (\tau) |^{p}  \leq
\big({\nu}_{p}(\Omega) \big)^{p} \,\, 
\sum \big( \Tr ( \theta \, X_{b} )\big)^{p}
\ee
Furthermore, from (\ref{defQC}) it follows that
\be\label{eqn6}
\Tr | \Phi(\theta) |^{p} =  \sum [\Tr (\theta X_{b})]^p 
\ee
Combining (\ref{eqn5}) and (\ref{eqn6}) and taking the $p^{\rm th}$ root gives
\be
|| \Omega \otimes \Phi (\tau) ||_{p}  \leq
{\nu}_{p}(\Omega) \,\, || \Phi(\theta) ||_{p}
\leq {\nu}_{p}(\Omega) \,\, {\nu}_{p}(\Phi) 
\ee
which then proves the result.

\bigskip
Turning now to the additivity of the channel capacity, we will again
establish the bound (\ref{suff1}). We claim that the following
identity holds:
\be\label{iden1}
S \bigg( (\Omega \otimes \Phi) (\tau) \,|\, {\rho}_{\Omega} \otimes 
{\rho}_{\Phi} \bigg) = 
\sum \, \Tr ( \theta \, X_{b} ) 
S \bigg( \Omega ({\tau}_{b}) \,|\, {\rho}_{\Omega} \bigg) +
S \bigg( \Phi (\theta) \,|\, {\rho}_{\Phi} \bigg)
\ee
From the result (\ref{SWthm}) it follows that
\be
S \bigg( \Phi (\theta) \,|\, {\rho}_{\Phi} \bigg) \leq
{\chi}^{*}(\Phi), \quad\quad
S \bigg( \Omega ({\tau}_{b}) \,|\, {\rho}_{\Omega} \bigg) \leq
{\chi}^{*}(\Omega)
\ee
Therefore (\ref{iden1}) implies
\be
S \bigg( (\Omega \otimes \Phi) (\tau) \,|\, {\rho}_{\Omega} \otimes 
{\rho}_{\Phi} \bigg) \leq
\sum \, \Tr ( \theta \, X_{b} ) {\chi}^{*}(\Omega) + {\chi}^{*}(\Phi) =
{\chi}^{*}(\Omega) + {\chi}^{*}(\Phi)
\ee
and this proves the result.

So it remains to verify the identity (\ref{iden1}). This follows
easily from the definition of relative entropy, and the fact that
$\{Q_{b}\}$ are orthogonal projections.

\section{Proofs for qubit channels}
In this section we prove Theorems \ref{thm2} and \ref{thm3}.
We do this by establishing the bound (\ref{new}), and then using
Corollary \ref{cor1}, which will be proved in the next section.

Let $\Phi$ be a qubit map, and assume that bases have been
chosen in its domain and range so that it has the form (\ref{Phi}).
Clearly, the maximal $l_p$ norm of $\Phi$ is
invariant under permutations of the three coordinates. It is also
invariant under the following symmetry operations.

\medskip
\begin{lemma}
For every $p$, ${\nu}_{p}(\Phi)$ is invariant if the signs of
any two of $({\lambda}_{1}, {\lambda}_{2}, {\lambda}_{3})$ are
reversed, or if the signs of any two of $(t_1, t_2, t_3)$ are
reversed.
\end{lemma}

The proof is easy: first notice that conjugation by ${\sigma}_1$ in
the domain of $\Phi$ switches the signs of ${\lambda}_{2}, {\lambda}_{3}$
without any other changes, and similarly for conjugation by
${\sigma}_2$ and ${\sigma}_3$. Then notice that simultaneous conjugation
by ${\sigma}_1$ in both the domain and range of $\Phi$ switches
the signs of $t_2, t_3$ without any other changes, and similarly for
${\sigma}_2$ and ${\sigma}_3$.

As a consequence, we will assume henceforth without loss of generality that
\be\label{assume}
t_1 \geq 0, t_2 \geq 0, \quad {\rm and} \quad 
{\lambda}_{1} \geq  {\lambda}_{2} \geq 0
\ee

\bigskip
Our first goal is to establish Conjecture \ref{conj1} for $p=2$, for any
map $\Phi$.
We rewrite (\ref{def:M}) more fully as
\be\label{def:M2}
M =  \left(\matrix{X & Y_1 - i Y_2 \cr Y_1 + i Y_2 & Z \cr}\right)
\ee
where $X > 0$, $Z > 0$ and $Y_1$, $Y_2$ are hermitian.
Let $W = (X + Z)/2$. Then
using the special form (\ref{Phi}) we get
\be\label{eqn:8}
& & (I \otimes \Phi) (M) = \\ \nonumber
& & \left(\matrix{c_{++} X + c_{-+} Z & 
(t_1 W + {\lambda}_{1} Y_1) - i (t_2 W + {\lambda}_{2} Y_2) \cr 
 & \cr
(t_1 W + {\lambda}_{1} Y_1) + i (t_2 W + {\lambda}_{2} Y_2) 
& c_{--} X + c_{+-} Z \cr}\right)  
\ee
where
\be\label{c's}
c_{++} = (1 + {\lambda}_{3} + t_3)/2, & \quad &
c_{-+} = (1 - {\lambda}_{3} + t_3)/2 \\ \nonumber
c_{+-} = (1 + {\lambda}_{3} - t_3)/2, & \quad &
c_{--} = (1 - {\lambda}_{3} - t_3)/2  
\ee
Note that since $M \geq 0$ and $\Phi$ is a qubit map, it follows that
$(I \otimes \Phi) (M) \geq 0$ for all choices of $X$ and $Z$.
Hence the four coefficients in
(\ref{c's}) are positive, for all allowed values of $t_3$ and ${\lambda}_3$.

We consider first the case that $p=2$, and $\Phi$ is any qubit map.
Taking the trace of the square of (\ref{eqn:8}) gives
\be
\Tr | (I \otimes \Phi) (M) |^2 & = &
\Tr (c_{++} X + c_{-+} Z)^2 + \Tr (c_{+-} X + c_{--} Z)^2 \\ \nonumber
& + & 2 \Tr (t_1 W + {\lambda}_{1} Y_1)^2 + 2 \Tr (t_2 W + {\lambda}_{2} Y_2)^2
\ee
Define
\be
x = || X ||_2, \quad
z = || Z ||_2, \quad
y_1 = || Y_1 ||_2, \quad
y_2 = || Y_2 ||_2
\ee
Then using the Cauchy-Schwarz inequality for the Hilbert-Schmidt norm, and 
our positivity condition (\ref{assume}) we get
\be\label{eqn:9}
\Tr | (I \otimes \Phi) (M) |^2 & \leq &
(c_{++} x + c_{-+} z)^2 + (c_{+-} x + c_{--} z)^2 \\ \nonumber
& + &
2 \bigg(t_1 {(x+z) \over 2} + {\lambda}_{1} y_1 \bigg)^2 + 
2 \bigg(t_2 {(x+z) \over 2} + {\lambda}_{2} y_1 \bigg)^2
\ee
Define the $2 \times 2$ matrix
\be
m = \left(\matrix{x & y_1 - i y_2 \cr y_1 + i y_2 & z \cr}\right)
\ee
Then (\ref{eqn:9}) can be re-written as
\be\label{eqn:10}
|| I \otimes \Phi(M) ||_{2} \leq
|| \Phi(m) ||_{2}
\ee
The positivity of $M$ implies that
\be
\Tr | Y_1 - i Y_2 |^2 = y_{1}^2 + y_{2}^2 \leq
x z,
\ee
and hence that $m$ is positive. Therefore
\be
|| (I \otimes \Phi) (M) ||_{2} \leq
{\nu}_{2}(\Phi) \Tr (m) =
{\nu}_{2}(\Phi) (x + z) =
{\nu}_{2}(\Phi) ( || X ||_2 + || Z ||_2)
\ee
which establishes ({\ref{new}) for $p=2$, and hence by Corollary \ref{cor1} proves
Theorem \ref{thm2}.

\bigskip
In order to prove Theorem \ref{thm3} we will 
assume that the condition (\ref{condition}) is satisfied.
Without loss of generality,
this condition can be rewritten as follows:
\be\label{assume2}
t_1 \geq 0 \quad {\rm and} \quad
t_2 = 0 \quad {\rm and} \quad 
{\lambda}_{1} \geq {\lambda}_{2} \geq 0.
\ee
To see this,
suppose first that $|{\lambda}_{i}| \neq |{\lambda}_{j}|$ for any
$i,j$. Then the condition (\ref{condition}) implies that at least one
of the $t_{i}$ is zero, and also that the corresponding $|{\lambda}_{i}|$
is not the largest. Hence by permuting coordinates we can arrange that
$t_2 = 0$ and that $|{\lambda}_{1}| > |{\lambda}_{2}|$. By switching signs
of pairs of parameters we can then re-state (\ref{condition}) as
(\ref{assume2}).
Suppose now that $|{\lambda}_{i}| = |{\lambda}_{j}|$ for some $i,j$.
By permuting coordinates we can assume that $|{\lambda}_{1}| = |{\lambda}_{2}|$,
and by changing signs that ${\lambda}_{1} = {\lambda}_{2} \geq 0$.
This allows a further symmetry transformation, namely we can conjugate
by a unitary matrix $U = e^{i \theta {\sigma}_{3}}$ in the range of
$\Phi$ without changing 
${\nu}_{p}(\Phi)$. With such a conjugation
we can set $t_2 = 0$, and then the condition (\ref{assume2}) again holds.

\medskip

The condition (\ref{assume2}) is clearly satisfied for all unital maps, since 
in that case 
$t_i = 0$ for all $i$. It is also satisfied by all maps in the
closure of the set of extreme points of the (convex) set of qubit maps. 
This fact follows
from Theorem 4 in \cite{RSW}, where it was shown that all such maps 
have only one of the parameters $t_1, t_2, t_3$ being non-zero.
\medskip

In order to prove (\ref{new}), we re-write (\ref{eqn:8}) as
\be\label{eqn:11}
(I \otimes \Phi) (M) & = & \left(\matrix{R_{11} & R_{12} \cr
R_{21} & R_{22} }\right) \\ \nonumber
&  & \\ \nonumber
& = &
R_{11} \otimes E_{11} +
R_{12} \otimes E_{12} + 
R_{21} \otimes E_{21} +
R_{22} \otimes E_{22}
\ee
where $E_{ij}$ is the $2 \times 2$ matrix with $1$ in position $(i,j)$ and
$0$ elsewhere, and where
\be\label{R's}
R_{11} & = & c_{++} X + c_{-+} Z, \\ \nonumber
R_{12} & = & (t_1 W + {\lambda}_{1} Y_1) - i  {\lambda}_{2} Y_2, \\ \nonumber
R_{21} & = & (t_1 W + {\lambda}_{1} Y_1) + i {\lambda}_{2} Y_2, \\ \nonumber
R_{22} & = & c_{--} X + c_{+-} Z
\ee
(we have used the condition (\ref{assume2}) to set $t_2=0$).

For integer $p$ we can evaluate $\Tr | (I \otimes \Phi) (M) |^p$ by
multiplying the right side of (\ref{eqn:11}) with itself $p$ times,
and taking the trace with respect to a product basis $e_i \otimes f_j$ where
$\{e_i \}$ span ${\bf C}^K$ and $f_1, f_2$ span ${\bf C}^2$.
The result is 
\be\label{eqn:12}
\Tr | (I \otimes \Phi) (M) |^p = 
\sum \Tr [ E_{i_1 j_1} E_{i_2 j_2} \dots E_{i_p j_p}] \,\,
\Tr [ R_{i_1 j_1} R_{i_2 j_2} \dots R_{i_p j_p}],
\ee
where the sum runs over all indices $i_1, j_1, \dots ,i_p, j_p = 1,2$. The coefficient
$\Tr [ E_{i_1 j_1} E_{i_2 j_2} \dots E_{i_p j_p} ]$ in each of these terms
is non-negative, since the matrices $\{E_{ij}\}$ are all non-negative.
Furthermore, repeated application of H\"older's inequality shows that
\be
|\Tr A_1 A_2 \dots A_p| \leq
||A_{1}||_{p} \,\, ||A_{2}||_{p} \, \dots \, ||A_{p}||_{p}
\ee
for any product of $p$ matrices.
Hence the sum in (\ref{eqn:12}) is bounded above by 
\be\label{eqn:13}
\Tr | (I \otimes \Phi) (M) |^p \leq
\sum \Tr [ E_{i_1 j_1} E_{i_2 j_2} \dots E_{i_p j_p}] \,\,
||R_{i_1 j_1}||_{p} ||R_{i_2 j_2}||_{p} \dots ||R_{i_p j_p}||_{p}
\ee

We define the $2 \times 2$ matrix
\be\label{def:m'}
m' = \left(\matrix{x' & y' \cr y' & z' \cr}\right)
\ee
where now
\be
x' = || X ||_{p}, \quad
z' = || Z ||_{p}, \quad
y' = || Y_1 - i Y_2 ||_{p}
\ee
The matrix $m'$ is positive. This can be seen most easily by noting that
the positivity of $M$ implies that $Y_1 - i Y_2 = \sqrt{X} \, T \,
\sqrt{Z}$ where $T$ is a contraction \cite{RSW}, and hence by H\"older's
inequality $y' \leq \sqrt{x' \, z'}$. Applying the map $\Phi$ gives
\be\label{Phi(m')}
\Phi(m') & = & [c_{++} x' + c_{-+} z'] \otimes E_{11} +
[t_1 (x' + z')/2 + {\lambda}_{1} y'] \otimes E_{12} \\ \nonumber
&  & +
[t_1 (x' + z')/2 + {\lambda}_{1} y'] \otimes E_{21} +
[c_{--} x' + c_{+-} z'] \otimes E_{22} 
\ee

Applying the same method to evaluate $\Tr | \Phi(m') |^p$ gives 
\be\label{eqn:14}
\Tr | \Phi(m') |^p = 
\sum \Tr [ E_{i_1 j_1} E_{i_2 j_2} \dots E_{i_p j_p}] \,\,
r_{i_1 j_1} r_{i_2 j_2} \dots r_{i_p j_p}
\ee
where
\be\label{def:r}
r_{11} & = & c_{++} \, x' + c_{-+} \, z', \\ \nonumber
r_{12} & = & r_{21} = t_1 \big(x' + z'\big)/2 + {\lambda}_{1} y', \\ \nonumber
r_{22} & = & c_{--} \, x' + c_{+-} \, z'
\ee

\medskip
We now claim that
\be\label{claim}
\Tr | (I \otimes \Phi) (M) |^p \leq \Tr | \Phi(m') |^p
\ee
If we assume for the moment that (\ref{claim}) is valid, then
it implies
\be
|| (I \otimes \Phi) (M) ||_{p} \leq || \Phi(m') ||_{p} \leq
{\nu}_{p}(\Phi) \,\, \Tr (m') \leq {\nu}_{p}(\Phi) \,\, (x' + z')
\ee
This proves (\ref{new}), which by Corollary \ref{cor1}
implies Theorem \ref{thm3}.

\bigskip
So it sufficient to demonstrate (\ref{claim}).
From (\ref{eqn:13}) and (\ref{eqn:14}) it is sufficient to show that
\be
||R_{ij}||_{p} \leq r_{ij}
\ee 
for all $i,j=1,2$. First, using the positivity of $c_{+ +}$ etc we have
\bee
||R_{11}||_{p}  & = & || c_{++} X + c_{-+} Z ||_{p} \leq 
c_{++} x' + c_{-+} z' = r_{11} \\ \nonumber
||R_{22}||_{p}  & = & || c_{+-} X + c_{--} Z ||_{p} \leq 
c_{+-} x' + c_{--} z' = r_{22}
\eee
The remaining bound also follows easily, since
\be\label{eqn:15}
|| R_{12} ||_{p} & = & || t_1 {(X + Z) \over 2} + 
{\lambda}_{1} Y_1 - i {\lambda}_{2} Y_2 ||_{p} \\ \nonumber
& \leq & || t_1 {(X + Z) \over 2}||_{p} +
|| ({\lambda}_{1} - {\lambda}_{2}) Y_1 + {\lambda}_{2}(Y_1 - i Y_2) ||_{p} \\ \nonumber
& \leq & t_1 (x' + z')/2 + ({\lambda}_{1} - {\lambda}_{2}) || Y_1 ||_{p}
+ {\lambda}_{2} || Y_1 -i Y_2 ||_{p} 
\ee
where in the last line we used (\ref{assume}).
Furthermore
\bee
|| Y_1 ||_{p} & = & || (Y_1 - i Y_{2})/2 + (Y_1 + i Y_{2})/2 ||_{p} \\
& \leq & {1 \over 2} || Y_1 - i Y_{2} ||_{p} + {1 \over 2} || Y_1 + i Y_{2} ||_{p} \\
& = & y'
\eee
Hence  (\ref{eqn:15}) becomes
\be
|| R_{12} ||_{p} \leq t_1 (x' + z')/2 + ({\lambda}_{1} - {\lambda}_{2}) y'
+ {\lambda}_{2} y' =  t_1 (x' + z')/2 + {\lambda}_{1} y' = r_{12}
\ee
which establishes the result.

\section{Proofs of Corollaries}
\subsection{Corollary \ref{cor1}}
Let $\Omega$ be any completely positive map on ${\cal B}({\bf C}^K)$,
and let $\tau$ be a state on ${\cal B}({\bf C}^K \otimes {\bf C}^2)$
of the form
\be\label{A,B,C}
\tau = \left(\matrix{A & B \cr B^{*} & C }\right)
\ee
where $A,B,C$ are $K \times K$ matrices, with $A \geq 0$, $C \geq 0$ and
$\Tr (A + C) = 1$.
Then $M = (\Omega \otimes I)(\rho)$ has the form (\ref{def:M}) with
$X = \Omega (A)$, $Y = \Omega (B)$ and $Z = \Omega (C)$. Hence from the definition of the
maximal $l_p$ norm it follows that
\be
|| X ||_{p} \leq {\nu}_{p}(\Omega)  \,\,  \Tr (A), \quad
|| Z ||_{p} \leq {\nu}_{p}(\Omega)  \,\,  \Tr (C)
\ee
Applying (\ref{new}) and using the facts that $(I \otimes \Phi) (M) =
(\Omega \otimes \Phi) (\rho)$ and 
\newline $\Tr (A) + \Tr (C) = \Tr (\rho) = 1$
we immediately deduce Corollary \ref{cor1}.
\medskip

\subsection{Corollary 2}

Recall that the entropy of a state $\rho$ is defined by
\be
S(\rho) = - \Tr \rho \, \log \rho
\ee
Using $\Tr \rho =1$ it follows that
\be
{d \over dp} \bigg( || \rho ||_{p}  \bigg)_{p=1} =
- S(\rho),
\ee
and hence that
\be\label{s_min}
{d \over dp} \bigg( {\nu}_{p}(\Phi)  \bigg)_{p=1}
= - S_{\rm min}(\Phi)
\ee
Therefore taking the derivative of (\ref{AHWconj}) at $p=1$ yields
immediately (\ref{add.Smin}).

\subsection{Corollary 3}
From the results of Section 2, it is sufficient to establish the
bound (\ref{suff1}). 
For any states $\omega$ and $\rho$ we have
\be
\log ( \omega \otimes \rho ) = \log \omega \otimes I
+ I \otimes \log \rho
\ee
Furthermore since $\Phi$ is a unital qubit map it follows that
its optimal average output state is
\be
{\rho}_{\Phi} = {1 \over 2} \, I
\ee
Since $\log ({1 \over 2} \, I) = - \log (2) I$ and $\Tr (\Omega \otimes \Phi) (\rho) = 1$ 
it follows that the left side of (\ref{suff1}) can be written as
\be\label{suff1'}
 - S\bigg((\Omega \otimes \Phi) (\tau)\bigg) - \Tr \bigg((\Omega \otimes \Phi)
(\tau) 
\, \log ({\rho}_{\Omega}) \otimes I\bigg) + \log (2)
\ee
Define
\be
\omega = {\Tr}_{2} \tau
\ee
Then the second term in (\ref{suff1'}) is equal to
\be
- \Tr \, \Omega (\omega) \, \log ({\rho}_{\Omega})
\ee
Also, the fact that $\Phi$ is unital implies that
\be\label{cap.Phi}
{\chi}^{*}(\Phi) = \log (2) - S_{\rm min} (\Phi)
\ee
Hence to prove (\ref{suff1}) it is sufficient to prove that
\be\label{suff2}
 - S((\Omega \otimes \Phi) (\tau)) 
- \Tr \, \Omega (\omega) \, \log ({\rho}_{\Omega})
\leq {\chi}^{*}(\Omega) - S_{\rm min}
(\Phi)
\ee

\medskip
Now we use the bound (\ref{new'}), which is implied by (\ref{new}).
Again let $\tau$ have the form (\ref{A,B,C}), so that
$M = (\Omega \otimes I) (\tau)$ has the form (\ref{def:M}) with
$X = \Omega(A)$ and $Z = \Omega(C)$. Let $a = \Tr A =
\Tr X$, and define the states
\be
\alpha = {1 \over \Tr A} \, A = {1 \over a} \, A, \quad\quad
\gamma = {1 \over \Tr C} \, C = {1 \over 1-a} \, C
\ee
Then using the notation of (\ref{def:xi,zeta}), $\xi = \Omega (\alpha)$
and $\zeta = \Omega (\gamma)$, and
(\ref{new'}) can be written
\be
S ((\Omega \otimes \Phi) (\tau) ) \geq S_{\rm min} (\Phi) + a S(\Omega(\alpha))
+ (1-a) S(\Omega(\gamma))
\ee
Comparing with (\ref{suff2}), it is sufficient to prove that
\be\label{suff3}
- a S(\Omega(\alpha)) - (1-a) S(\Omega(\gamma)) 
- \Tr \, \Omega (\omega) \, \log ({\rho}_{\Omega})
\leq {\chi}^{*}(\Omega)
\ee
Since $\omega = a \alpha + (1-a) \gamma$, we can rewrite the left side of
(\ref{suff3}) as
\be\label{suff4}
a S \bigg(\Omega(\alpha) \,|\, {\rho}_{\Omega} \bigg)
+ (1-a) S \bigg(\Omega(\gamma) \,|\, {\rho}_{\Omega} \bigg)
\ee
Since ${\rho}_{\Omega}$ is the optimal output state for the channel $\Omega$, 
it follows from (\ref{OPWthm}) that
\be\label{last}
S \bigg(\Omega(\alpha) \,|\, {\rho}_{\Omega} \bigg)
& \leq & {\chi}^{*}(\Omega) \\
S \bigg(\Omega(\gamma) \,|\, {\rho}_{\Omega} \bigg)
& \leq & {\chi}^{*}(\Omega)
\ee
Combining (\ref{suff3}), (\ref{suff4}) and (\ref{last}) yields the result.

\section{Summary}
The results in this paper all concern product channels of the form
$\Omega \otimes \Phi$, where in every case $\Omega$ is an arbitrary
channel. For these product channels we prove a variety of results
involving different measures of the purity of output states from the channel.

The first set of results apply when $\Phi$ is a CQ or QC channel.
Recall that the CQ channel first maps an input state to a letter in a classical alphabet, and then
maps this to a quantum state at the output. The QC channel measures
the input state with some POVM, and assigns different results to orthogonal
output states. In both cases we prove that the output state with maximal
$l_p$ norm is a product state, and also that the Holevo capacity is achieved 
on a product state. In other words, the maximal $l_p$ norm of the product
channel is multiplicative and the Holevo capacity is additive.
These results were previously shown to be true in the case where $\Phi$ is the identity map
(and the additivity of the Holevo capacity also when $\Omega = \Phi$).

The second set of results apply when $\Phi$ is a qubit map, that is a map
on states in ${\bf C}^2$. We prove multiplicativity for the $p=2$ norm,
for any qubit map $\Phi$. We also prove multiplicativity for the $l_p$ norm
when $p$ is any integer, and with some restrictions on $\Phi$. The class
of maps $\Phi$ satisfying the restrictions includes all unital qubit maps.

The third set of results revolves around a conjectured bound (\ref{new}) 
for the $l_p$ norm
of any output state from the half-noisy channel $I \otimes \Phi$,
when $\Phi$ is a qubit channel. We show that this bound implies 
multiplicativity of the $l_p$ norm for any product channel $\Omega \otimes \Phi$.
We also show that when $\Phi$ is {\it unital} the bound implies
additivity of the Holevo capacity of the product channel $\Omega \otimes \Phi$.
Therefore we believe that this conjecture provides a new and useful approach to
the conjecture that the Holevo capacity is universally additive.
In a hopeful sign of future progress on this important problem,
Lieb and Ruskai have established Conjecture \ref{conj1} in one non-trivial case
(their proof appears as the Appendix below).

\bigskip

\noindent {\bf Acknowledgment:}  
The author thanks M. B. Ruskai for useful discussions and comments.
The author is also grateful to E. H. Lieb and M. B. Ruskai for allowing
their proof of a special case of Conjecture \ref{conj1}, eq. (\ref{new}) to 
appear as an Appendix to this paper.
This research was supported in part by 
National Science Foundation Grant DMS-97-05779.

\appendix

\section{Appendix: Theorem of Lieb and Ruskai}

Let $M = \pmatrix{X & Y \cr Y^* & Z}$ and recall that
$M$ is positive semi-definite if and only $Y = \sqrt{X} R \sqrt{Z}$
with $R$ a contraction.  Moreover, any contraction can be written
as a convex combination of unitary matrices.  (See \cite{HJ2} or
\cite{RSW} for details and further references.)  Hence, by
the  convexity of the $p$-norm, it suffices to prove (\ref{LR}) under
the assumption that $Y = \sqrt{X} V \sqrt{Z}$ with $V$ unitary.

We now consider the special case $X = Z$ and note that
we can write
\be
(I \ot \Phi)(M) = \pmatrix{X & \lambda Y \cr \lambda Y^* & X} 
=  \sqrt{F} G \sqrt{F}
\ee
with $F = \pmatrix{X & 0 \cr 0 & X} $ and
    $G =    \pmatrix{ I & \lambda V \cr \lambda V^* &I} $.  We 
will use a result of Lieb and Thirring  (Appendix B of \cite{LT}) that,
for $p
\geq 1$ and $F, G \geq 0$,
\be
    \tr (F^{1/2} G F^{1/2})^p \leq \tr (F^p G^p ).
\ee
The critical feature is to note that $G$ has eigenvalues 
$(1 \pm \lambda)$.  Moreover, 
\be \label{diag}
 \pmatrix{ I & \lambda V \cr \lambda V^* &I} = \half
  \pmatrix{ I &  V \cr  V^* & -I} 
    \pmatrix{ (1 + \lambda )I & 0 \cr 0 & (1 - \lambda )I}
    \pmatrix{ I &  V \cr  V^* & -I} 
\ee
Thus
\bee
\lefteqn {\tr [(I \ot \Phi)(M)]^p} \\ & \leq &   \half \tr 
    \pmatrix{ I &  V \cr  V^* & -I} \pmatrix{X^p & 0 \cr 0 & X^p} 
    \pmatrix{ I &  V \cr  V^* & -I} 
        \pmatrix{ (1 + \lambda )^pI & 0 \cr 0 & (1 - \lambda )^pI} \\
  & = & (1 + \lambda )^p \, \tr  \half \big( X^p +  V X^p V^*)  + 
   (1 - \lambda )^p  \, \tr  \half \big( X^p +  V^* X^p V)\\
 & = &   [2 \nu_p(\Phi)]^p  \,  \| X \|_p^{p} .
\eee
Taking the p-th root gives the desired result,  
$\| (I \ot \Phi)(M) \|_p \leq \nu_p(\Phi) \, 2 \| X \|_p$.

{~~}

\end{document}